\documentclass{article}

\usepackage{arxiv}

\usepackage[utf8]{inputenc} % allow utf-8 input
\usepackage[T1]{fontenc}    % use 8-bit T1 fonts
\usepackage{hyperref}       % hyperlinks
\usepackage{url}            % simple URL typesetting
\usepackage{booktabs}       % professional-quality tables
\usepackage{amsfonts}       % blackboard math symbols
\usepackage{nicefrac}       % compact symbols for 1/2, etc.
\usepackage{microtype}      % microtypography
\usepackage{lipsum}
\usepackage{graphicx}
\usepackage{amsmath}
\usepackage{multirow}
\graphicspath{ {./images/} }
\usepackage{caption}
\title{Unsupervised and Supervised Algorithms for Identification of Sample Pixels in FTIR Images}

\author{
 Xiangyu Zhao \\
  School of Biomedical Engineering\\
  Shanghai Jiao Tong University\\
  Shanghai 200240, China \\
  \texttt{xiangyu\_zhao@sjtu.edu.cn} \\
  \And
  Yudong Tian \\
  School of Biomedical Engineering\\
  Shanghai Jiao Tong University\\
  Shanghai 200240, China \\
  \texttt{jackson1@sjtu.edu.cn} \\
  \And
  Jingzhu Shao \\
  School of Biomedical Engineering\\
  Shanghai Jiao Tong University\\
  Shanghai 200240, China \\
  \texttt{answer000@sjtu.edu.cn} \\
   \And
 Chongzhao Wu \thanks{Corresponding Author}\\
  School of Biomedical Engineering\\
  Shanghai Jiao Tong University\\
  Shanghai 200240, China \\
  \texttt{czwu@sjtu.edu.cn} \\
}

\begin{document}
\maketitle
\begin{abstract}
Mid-InfraRed spectroscopy is a promising label-free technique that can offer insights into morphological and pathological alterations in biological tissues at the molecular level. Owing to the development of the Fourier Transform InfraRed (FTIR) spectrometer, combined with scanning devices, FTIR images can be produced by simultaneously acquiring spectral data from multiple spatial points, generating comprehensive chemical maps. In the data pre-processing, the identification of the sample pixels, with the background pixels excluded, is important for further effective feature extraction in FTIR images. Here, we present three algorithms realized in unsupervised and supervised approaches for the identification of the sample pixels. The algorithms demonstrate accurate prediction results of the sample and background pixels, and the supervised method further enables the automatic detection. These findings highlight thorough and robust solutions to the sample pixels detection problem in FTIR images, contributing to the FTIR signal processing and future research with FTIR images.
\end{abstract}

% keywords can be removed
%\keywords{First keyword \and Second keyword \and More}

\section{Introduction}
Mid-InfraRed (MIR) spectroscopy appears as a pivotal field of research concerning the molecular characterization. Based on the absorption of MIR light by various chemical bonds, such as S-H, C-H, N-H, and O-H, this technology has found widespread and diverse applications \cite{babu2024fuel,deng2024evaluation,zhu2024surface} driven by significant advances in instrumentation and analytical algorithms \cite{pilling2016fundamental}. In biomedical and clinical research, MIR spectroscopy has proven to be exceptionally valuable due to its ability to detect natural tissue biomarkers such as proteins, lipids, nucleic acids, and carbohydrates, making it a promising pathological tool.

Owing to the development of Fourier Transform InfraRed (FTIR) spectroscopy, the spectrometer can be designed to combine with scanning devices such as motorized translation stages and array detectors, enabling tissue microspectroscopic imaging by MIR absorbance spectroscopy. Building upon this foundation, FTIR imaging extends the capabilities of MIR spectroscopy by enabling the simultaneous acquisition of spectral data from multiple spatial points, generating comprehensive chemical maps that reveal the distribution of chemical species across a sample surface. This integration of spectral and spatial information has revolutionized the analysis of complex systems, allowing researchers to investigate microstructural heterogeneity and molecular interactions with unprecedented detail. As a result, FTIR imaging has become an indispensable technique in modern analytical science, driving advances in both fundamental research and applied technologies. When FTIR imaging is applied in clinical research, the histological structures can be identified without any labeling or staining agent. This approach, named spectral histopathology (SHP) \cite{boutegrabet2021automatic}, has greatly enhanced the understanding of diseases affecting various organs, including brain \cite{bergner2013tumor}, breast \cite{verdonck2016characterization,ali2019investigation,ali2018simple},  colon \cite{nallala2020characterization,song2020micro}, kidney \cite{shao2024label,varma2016label}, liver \cite{liyanage2020fourier}, lung \cite{yang2021diagnosis,li2024comparison}, lymphomas \cite{lindtner2023comparison},  ovarian \cite{li2018characterization}, skin \cite{zhang2019investigation}, and thyroid \cite{da2024expression}.

Existing imaging technologies and computer storage formats necessitate that FTIR images are typically stored in rectangular grids. However, both tissue sections and cellular samples often contain regions of interest (ROIs) with irregular morphologies. As a result, when acquiring complete FTIR images of a tissue sample, non-tissue regions are inevitably included, leading to background pixels that contain MIR spectra from non-biological materials. This is particularly relevant in clinical and pathological studies, where a significant proportion of biopsy samples are processed and preserved as formalin-fixed paraffin-embedded (FFPE) tissues—a standard procedure for long-term storage and histological analysis \cite{boutegrabet2021automatic,ly2008combination}. The paraffin matrix, along with other substrates or contaminants, contributes strong infrared absorption features that can dominate the spectral profile and are non-relevant to pathological information, thereby obscuring or distorting the underlying biochemical spectral signals. This interference poses a significant challenge to accurate spectral interpretation and reduces the reliability of chemical mapping. Consequently, these background spectra introduce noise and complicate data analysis, calling for robust computational strategies to effectively detect, isolate, and remove them, and finally identify the sample pixels for further analysis.

In conventional detection, single-band absorption-based methods are commonly employed to identify spectra of the sample pixels and remove background pixels. For instance, background spectra are usually excluded based on the absence of absorption peaks in the Amide I/II region (1700–1500 $cm^{-1}$). However, variations in sample thickness can introduce baseline absorbance across the entire spectral range and spatial points, even in background pixels, and baseline drift is also frequently observed. Moreover, the distribution of absorption peaks varies among different types of samples, which may introduce misleading results or prevent the use of a unified absorption value criterion across all samples, thereby reducing the automation level of the algorithm. Consequently, it is necessary to adopt methods that consider the full spectral range to improve the accuracy and robustness of sample pixels identification and background removal.

\begin{figure}[!htp]
    \centering
    \includegraphics[width=\linewidth]{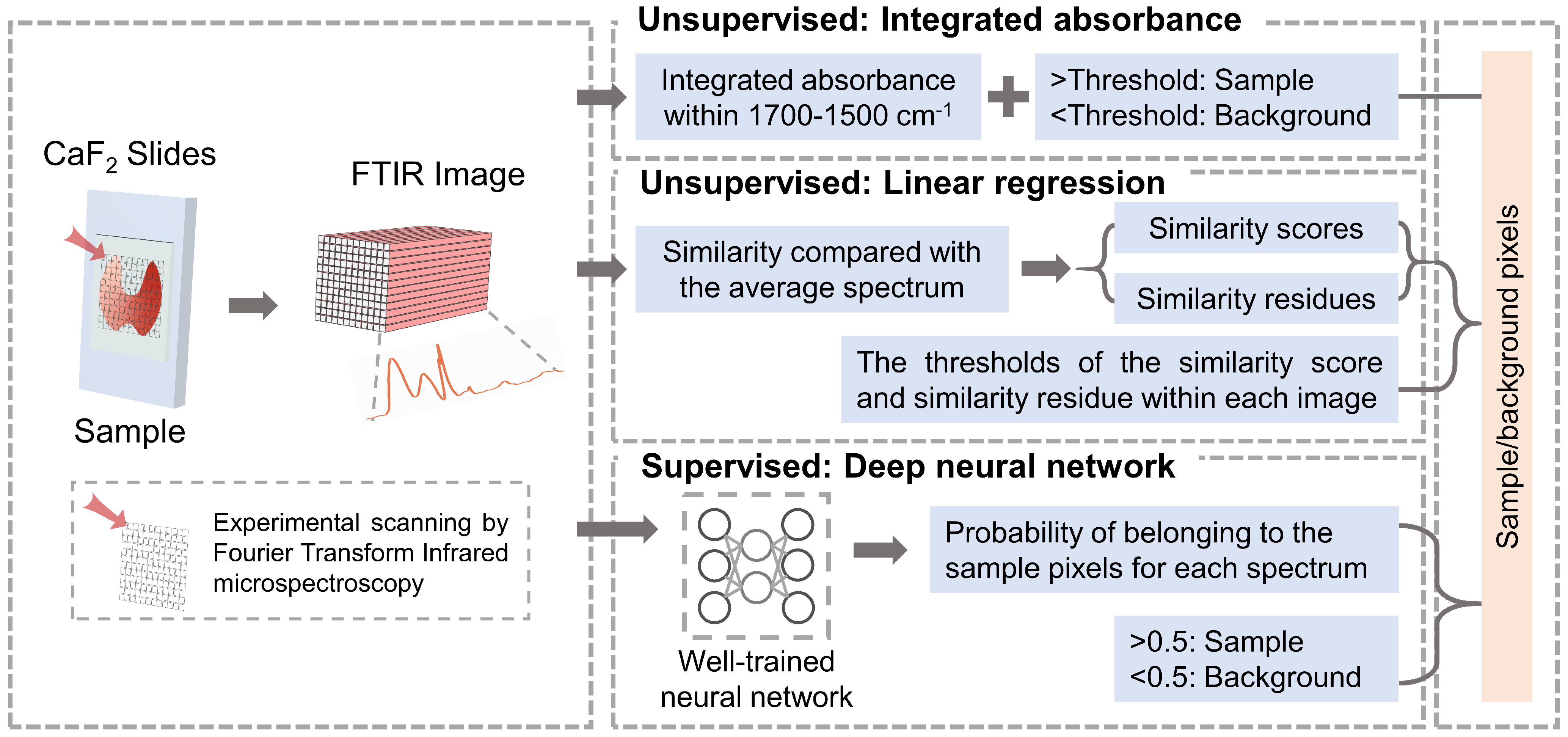}
    \caption{\textbf{Workflow of the multivariate sample detection algorithms for FTIR images.} The algorithms are realized with two approaches in this work: supervised and unsupervised frameworks. In the unsupervised approach based on integrated absorbance, the integrated absorbance of pixels is calculated between 1700-1500 cm$^{-1}$ and the sample pixels are predicted with a selected threshold. In the unsupervised approach based on linear regression, the algorithm first computes similarity scores and similarity residues of all spectra compared to the sample’s average spectrum. Dynamic thresholds of the similarity score and similarity residue for the detection of the sample pixels are then determined based on the specific characteristics and criteria of each sample. In the supervised approach, the spectrum of each pixel is fed into a well-trained neural network to predict its probability of belonging to the sample pixels. Pixels with predicted probabilities below 0.5 are automatically classified as background pixels, with others classified as sample pixels.
}
    \label{fig:Workflow}
\end{figure}

In this work, we proposed multiple algorithms for sample pixel identification in FTIR images and offer a comparison. As shown in Figure \ref{fig:Workflow}, the algorithms are realized with three approaches in this work: an unsupervised framework based on integrated absorbance, an unsupervised framework based on linear regression, and a supervised framework based on a deep neural network (DNN). In the unsupervised methodology based on integrated absorbance, the integrated absorbance is calculated between 1700-1500 cm$^{-1}$, and the sample pixels are selected by comparing with the threshold. In the unsupervised methodology based on linear regression, the algorithm computes similarity scores and similarity residues of all spectra compared to the sample’s average spectrum by fitting them to a linear model. The final detections are all based on the scores and residues, which are calculated without the guidance of any label. In the supervised methodology, the spectrum of each pixel is fed into a well-trained neural network with labeled sample spectra to predict its probability of belonging to the sample pixels and background pixels. Given the end-to-end structure of the deep neural network, the supervised approach enables direct spectral-to-classification prediction without the need for manual thresholding, facilitating the automated identification. Furthermore, to confirm the universality, our algorithm
 has been evaluated on infrared images tested with tissue sections of different thicknesses from multiple organs in this paper. By offering unsupervised and supervised algorithms, this paper presents thorough and robust solutions to the sample pixels detection problem in FTIR images, which will contribute to the FTIR signal processing and future research with FTIR images.

\section{Materials and Methods}
\subsection{Sample preparation}
The samples are collected from human thyroids and mouse kidneys in the form of tissue blocks. For samples of human thyroids, patients were enrolled at the First Hospital of the China Medical University, approved by the Ethics Committee of the First Hospital of China Medical University (Number 2024-582). For samples of rat kidneys, the female specific pathogen-free (SPF) Sprague Dawley rats, weighing 200 g to 240 g, were obtained from Chang Zhou Cavens Laboratory Animal Ltd (China) and were raised under identical conditions and sacrificed after two weeks for kidney samples, approved by the Institutional Animal Care and Use Committee at Jiangsu Science Standard Medical Testing Company Limited (IACUC22-0107). Tissue blocks obtained during surgical procedures were fixed in formalin and embedded in paraffin (FFPE).

For each FFPE block, two consecutive tissue sections were prepared by using a microtome. The first section was mounted on a calcium fluoride (CaF$_2$) slide to collect the ground truth spectral data. The second section was mounted on a glass slide and stained with hematoxylin and eosin (H\&E), followed by scanning with a digital slide scanner (Leica Aperio CS$_2$) for pathological examination. The tissue sections of the thyroid were prepared with a thickness of 4 $\mu m$, while those of the kidney were prepared with a thickness of 2 $\mu m$. Samples with different thicknesses can help demonstrate the universality of the proposed algorithms.

\subsection{Collection of infrared spectroscopic images}

The MIR spectra were collected from FFPE CaF$_2$ slices in transmission mode. Spectral data were experimentally measured by a Fourier-transform infrared spectrometer (Bruker Vertex 80v) coupled to a microscope (Hyperion 2000) equipped with a liquid nitrogen cooled MCT pointed detector in transmission mode and a software-controlled x-y stage. Broadband Globar served as the radiation source. FTIR mapping were performed in the spectral range from 4000 cm$^{-1}$ to 600 cm$^{-1}$ with the spectral resolution of 4 cm$^{-1}$. The so-called fingerprint region used in our algorithm targets from the wavenumber of 1800 cm$^{-1}$ to 1000 cm$^{-1}$.

To map the entire tissue slices, FTIR spectra were collected with the built-in Globar source with a 15$\times$ objective. Pixels were collected for each tissue section at a step size of 100 $\mu m$ in both x and y directions with a $100\times100$ $\mu m^2$ square aperture defined by knife edges. 8 scans were averaged per pixel spectrum, and the background spectra (32 scans were averaged) were recorded per collection of 50 pixels.

\subsection{Integrated absorbance-based unsupervised detection}

When it is assumed that the background pixels consist solely of the substrate and paraffin pixels, then the absence of absorption peaks in the Amide I/II region can be exploited to screen sample spectra against the background spectra by the integrated absorbance in this spectral region. To validate the effectiveness of this method in the screening of sample pixels, and to compare it with the two subsequent multivariate approaches, integrated absorbance of each pixel is calculated with the composite trapezoidal rule in 1700-1500 cm$^{-1}$.

By scanning the threshold of integrated absorbance $T_{Abs.}$ and calculating the predicted map of samples with Equation \ref{equ:MapInt}, we are able to obtain an identification result that is most consistent with the roughly labeled maps of each FTIR image. We use the Jaccard index between the prediction maps and the labeled ground truth maps in Equation \ref{equ:JI} to quantify the performance:

\begin{equation}
Map_{Sample}(x,y) = \left\{\begin{matrix}
1 & a(x,y)>T_{Abs.}\\
0 & else
\end{matrix}\right.
\label{equ:MapInt}
\end{equation}

\begin{equation}
Jaccard\; index = \frac{card(Map_{Prediction} \cap Map_{Ground\; truth})}{card(Map_{Prediction} \cup Map_{Ground\; truth})}
\label{equ:JI}
\end{equation}

where $card(Map)$ represents the number of pixels that are valued with 1 in the proposed $Map$.

\subsection{Linear regression-based unsupervised similarity analysis}

In the unsupervised method, the identification of sample spectra primarily relies on the position of pixel spectra within the overall spectral distribution, representing a spectral distribution-based approach. This method mainly employs linear regression to compute the similarity of each spectrum compared to the average spectrum (center of distribution). The multivariate linear regression model utilized in this algorithm has also been previously applied in the extended multiplicative signal correction (EMSC) algorithm \cite{afseth2012extended}. In such a linear model, each absorbance spectrum $A$ can be viewed as a linear combination of the components as shown in Equation \ref{equ:linearregression}.

\begin{equation}
A(\nu) = a\cdot \bar{A}(\nu) + \textit{\textbf{b}}\cdot \textit{\textbf{P}}(\nu) + e(\nu)
\label{equ:linearregression}
\end{equation}

\begin{equation}
\textit{\textbf{b}}\cdot \textit{\textbf{P}}(\nu) = b_0 + b_1 \cdot \nu + b_2 \cdot \nu^2 +b_3 \cdot \nu^3 + b_4 \cdot \nu^4
\label{equ:polynomial}
\end{equation}

\begin{equation}
    E = \ln(\sum^{\nu_{max}}_{\nu=\nu_{min}} e(\nu)^2 )
    \label{equ:fittingerror}
\end{equation}
where $\nu$ represents the wavenumber and $A(\nu)$ represents the absorbance of the current wavenumber $\nu$. $\bar{A}(\nu)$ is the average absorbance of wavenumber $\nu$ within the image. $\textit{\textbf{P}}(\nu)$ is a fourth-order polynomial of wavenumbers used to model the baseline and light scattering effect, with the coefficient $\textit{\textbf{b}}=[b_0, b_1, ..., b_4]$, as depicted in Equation \ref{equ:polynomial}. $e(\nu)$ is the modeling error of wavenumber $\nu$. In each spectrum, the absorbance $A(\nu)$ for all wavenumbers shares the same coefficients
 $a$ and $\textit{\textbf{b}}$, which are calculated by linear regression based on Equation \ref{equ:linearregression}. The fitting error $E$ for each spectrum is calculated with Equation \ref{equ:fittingerror}, where the natural logarithm is used to compress the value distribution.

Based on the above definitions, we aim to extract the sample pixels by evaluating the similarity between the spectrum of each pixel in the FTIR image and the average spectrum of the image. To this end, we further define the fitted coefficient $a$ as the \textbf{similarity score} of the spectrum compared with the average spectrum, which serves as the primary criterion for determining whether a pixel spectrum corresponds to the background spectrum. We also define $E$ as the \textbf{similarity residue} of the spectrum compared with the average spectrum, which provides auxiliary inference in certain cases. The final map of sample pixels is generated with the given thresholds for similarity score and similarity residue, termed $T_{Score}$ and $T_{Residue}$, by taking the intersection between the sets $a > T_{Score}$ and $E<T_{Residue}$ as shown in Equation \ref{equ:MapGenerate}:
\begin{equation}
Map_{Sample}(x,y) = \left\{\begin{matrix}
1 & a(x,y)>T_{Score}\; and\; E(x,y)<T_{Residue} \\
0 & else
\end{matrix}\right.
\label{equ:MapGenerate}
\end{equation}

Based on these two definitions related to similarity, we will proceed with the analysis for the subsequent extraction of the background pixels. With the proper selection of thresholds of the similarity score and residue, detection of the background pixels can be fulfilled with different criteria concerning the spectrum distribution. In this work, to give an example, we change the thresholds to fit the prediction results to the roughly labeled maps of each FTIR image, which are evaluated with the Jaccard index as Equation \ref{equ:JI}.

\subsection{Deep neural network-based supervised automatic classification}

\begin{figure}[!htp]
    \centering
    \includegraphics[width=\linewidth]{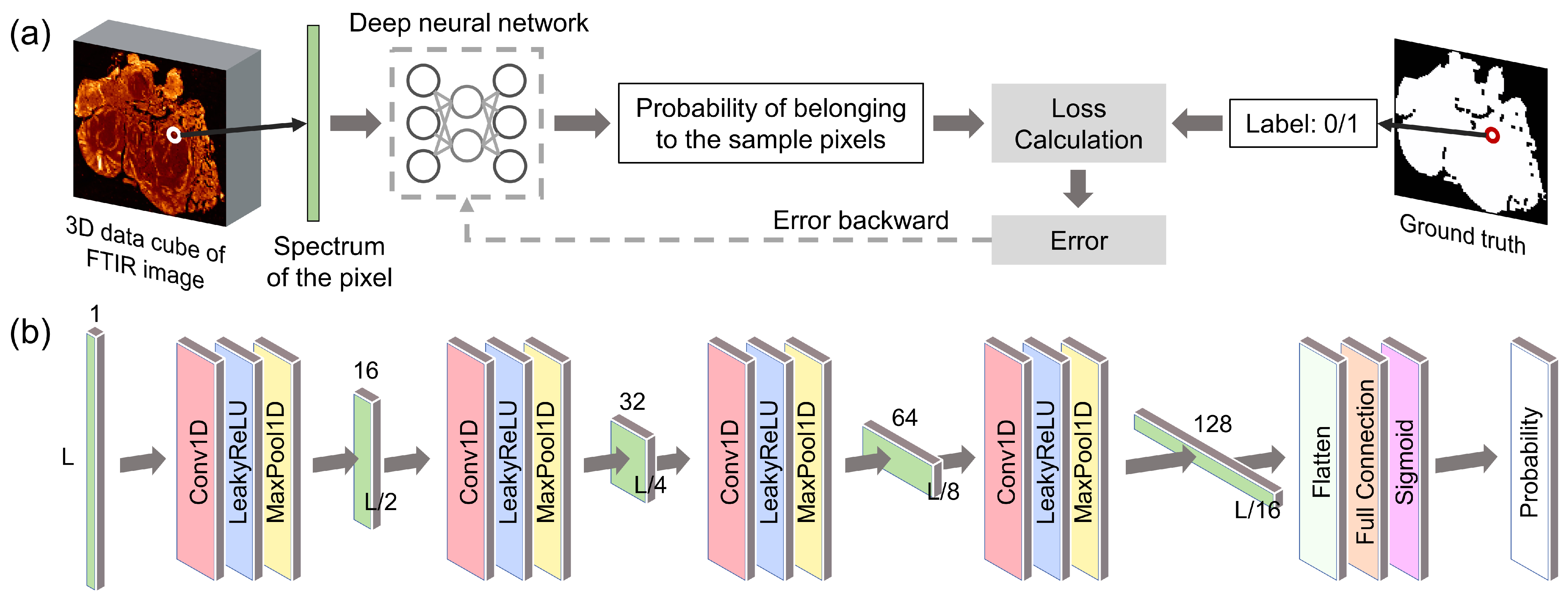}
    \caption{\textbf{Details of the deep neural network (DNN) for automatic sample pixels detection in FTIR images.} (a) The training procedure of the DNN. The spectrum and label of each pixel are paired as an 'input-ground truth' issue. The DNN calculates the probability of the pixel belonging to the sample pixels based on the input spectrum, which is finally compared with the ground truth label for the error and its backward. (b) The backbone structure of the DNN consists of four 1D convolution layers and one full connection layer for calculating the probability from the input spectrum. The data tensors are represented with the green rectangles with the shape of Channels$\times$Lengths.}
    \label{fig:DNNStructure}
\end{figure}

Considering that the unsupervised method requires threshold customization for each sample, they exhibit rather lower stability and robustness to interference, which further hinders the automation of the overall sample and background pixels detection process. In contrast, deep neural network models can directly provide classification results for each spectrum through an end-to-end architecture, enabling unified and automated detection of background spectra with all wavenumber channels considered. The backbone deep neural network architecture and its training are detailed in Figure \ref{fig:DNNStructure}.

 The deep neural network consists of four convolution layers and one full connection layer. Each convolution layer contains a 1$\times$3 convolution kernel, a leaky rectified linear (LReLU) activation function to 
avoid vanishing gradient, and a maxpooling layer. The LReLU is defined as:

\begin{equation}
    LReLU(x)=\left\{\begin{matrix} x & x\geq 0 \\ 0.2x & x<0 \end{matrix}\right.
    \label{equ:LReLU}
\end{equation}

Specifically, the network input comprises one channel, consisting of the sample pixel spectrum shaped $1\times416$. After being processed with the deep neural network, a $1\times 1$ probability would be calculated as an output. Finally, the probability provided by the DNN is compared with the ground truth label with Mean Squared Error (MSE) as a loss function.

Additionally, during the training procedure, we implement a learning rate schedule, gradually decreasing from  10$^{-3}$ to 10$^{-5}$ over 500 epochs. The batch size was set at 4,096 to facilitate the learning of the model across multiple samples, increase stability, and prevent overfitting. The DNN model is implemented using Python version 3.10.14 and Pytorch version 2.2.0. The training is performed on a desktop computer with an Intel Xeon W-2235 central processing unit (CPU), 64 GB of random-access memory (RAM), and a Nvidia GeForce GTX 3090 graphics processing unit (GPU). It takes about 0.0006s to handle a single spectrum through the proposed DNN.

\section{Results and Discussion}

\subsection{Collection and IR measurements of the tissue section samples}

Ultimately, we collected a total of 4 human thyroid tissue sections from patients and 3 rat kidney tissue sections from rats and tested as discussed in Materials and Methods. The sample pixel regions and background spectral regions were roughly annotated, as shown in Figures \ref{fig:UnsupervisedThyroid} and \ref{fig:UnsupervisedKidney}. In the unsupervised approach, each sample is analyzed independently, and for each sample, the similarity score and similarity residue between every pixel spectrum and the average spectrum within that sample are computed. While in the supervised DNN approach, Samples 1–3 and Samples 5–6 were used as the training set to train the model, while Samples 4 and 7 were used as the test set to evaluate model performance, as shown in Table \ref{tab:sample}.

\begin{table}[!htp]
\centering
\caption{\textbf{Sample preparation and spectral dataset information}}
\label{tab:sample}
\begin{tabular}{ccccc}
\hline
Sample ID & Source                         & Image size     & Number of spectra & Group for DNN Training \\ \hline
1         & \multirow{4}{*}{Human Thyroid} & 103$\times$109 & 11,227            & Train         \\
2         &                                & 120$\times$138 & 16,560            & Train         \\
3         &                                & 116$\times$134 & 15,544            & Train         \\
4         &                                & 113$\times$114 & 12,882            & Test          \\ \hline
5         & \multirow{3}{*}{Rat Kidney}    & 124$\times$136 & 16,864            & Train         \\
6         &                                & 118$\times$139 & 16,402            & Train         \\
7         &                                & 125$\times$142 & 17,750            & Test          \\ \hline
\end{tabular}
\end{table}

\subsection{Results of the detection based on the integrated absorbance}

\begin{figure}[!htp]
    \centering
    \includegraphics[width=\linewidth]{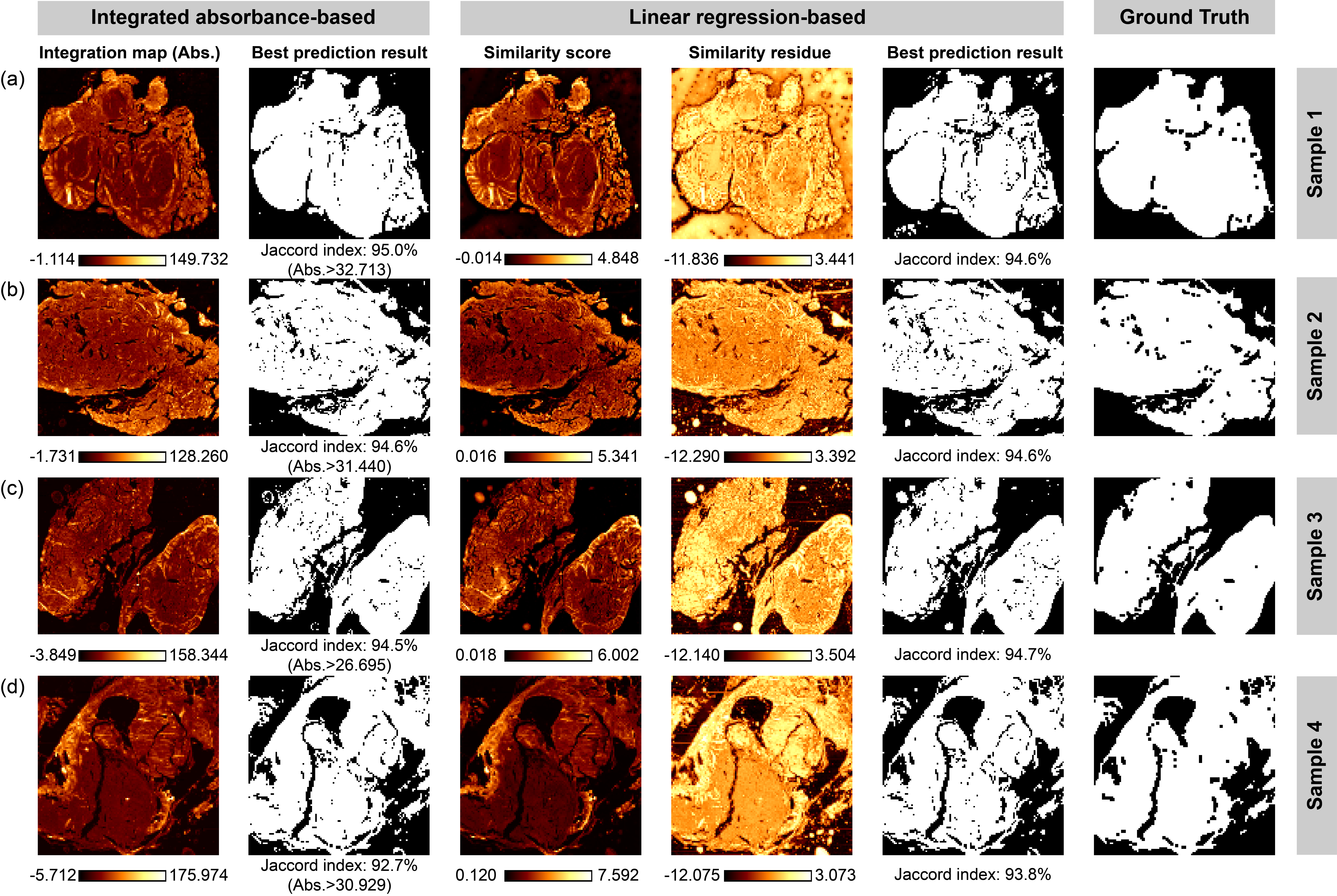}
    \caption{\textbf{Results of background detection with unsupervised linear regression approach on human thyroid tissue section samples.} The integrated absorbance-based results of each sample are shown with a map of integrated absorbance (1700-1500 $cm^{-1}$), and a best prediction map. The linear regression-based results are shown with a map of similarity score, a map of similarity residue, and a best prediction map. A ground truth detection map of each sample is also shown. In the ground truth map and the best prediction map, the white-colored pixels are sample pixels, and the black ones are background. The samples from the thyroid are labeled with 'Sample 1-4' and shown in (a-d), respectively.}
    \label{fig:UnsupervisedThyroid}
\end{figure}

\begin{figure}[!htp]
    \centering
    \includegraphics[width=\linewidth]{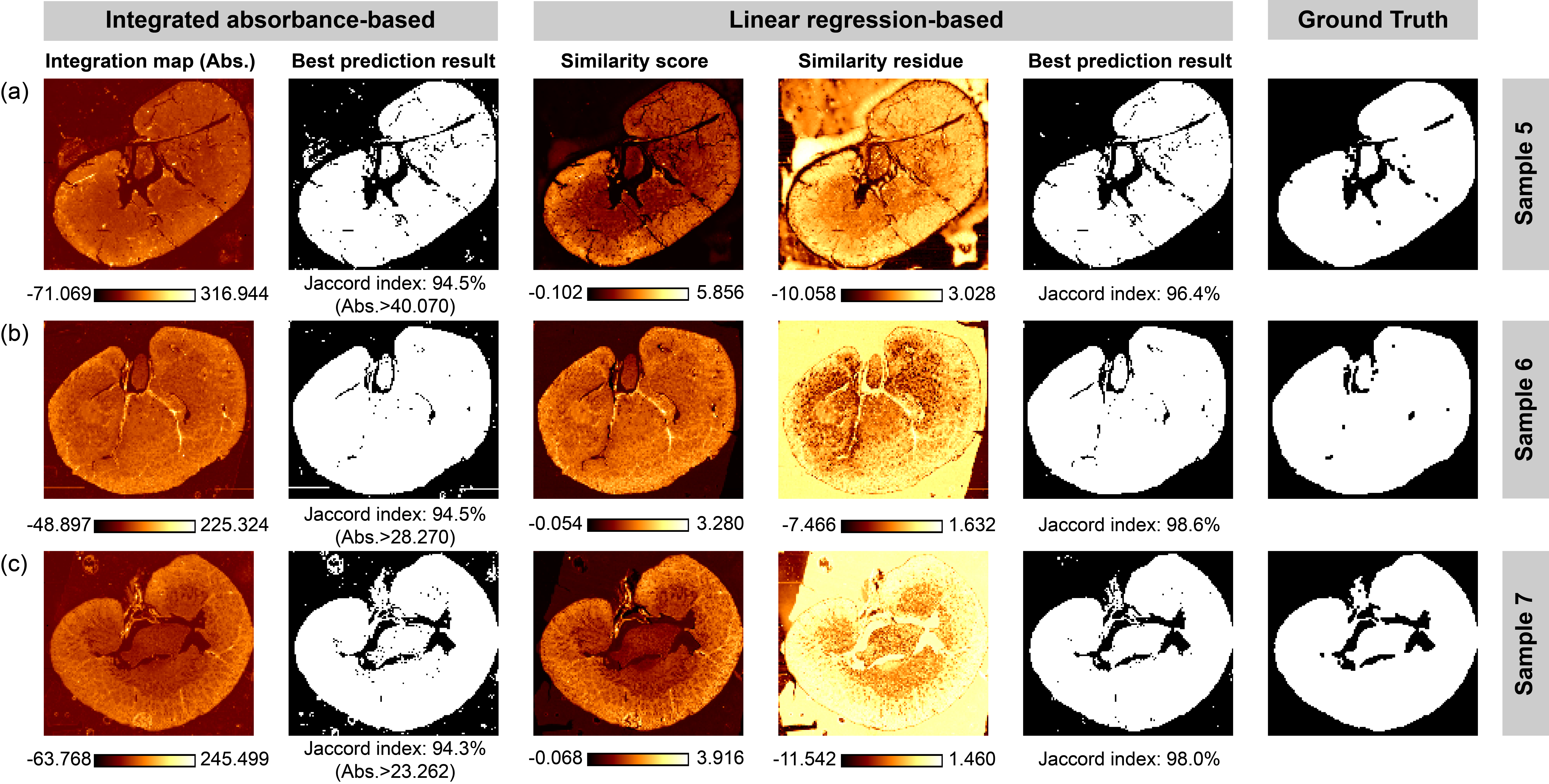}
    \caption{\textbf{Results of background detection with unsupervised linear regression approach on mouse kidney tissue section samples.} The integrated absorbance-based results of each sample are shown with a map of integrated absorbance (1700-1500 $cm^{-1}$), and a best prediction map. The linear regression-based results are shown with a map of similarity score, a map of similarity residue, and a best prediction map. A ground truth detection map of each sample is also shown. In the ground truth map and the best prediction map, the white-colored pixels are sample pixels, and the black ones are background. The samples from the thyroid are labeled with 'Sample 5-7' and shown in (a-c), respectively.}
    \label{fig:UnsupervisedKidney}
\end{figure}

After identifying the sample pixels based on the integration results, the outcomes are shown in Figures \ref{fig:UnsupervisedThyroid} and \ref{fig:UnsupervisedKidney}. The human thyroid sample was well-prepared and exhibited high-quality spectral measurements, with minimal contamination pixels whose spectra overlap with the sample spectra in absorption peaks, and relatively little baseline drift in the background pixels. As a result, the integration-based method achieved identification results of comparable quality to the other unsupervised multivariate approach. However, in the rat kidney samples, a higher number of spectral points exhibited background drift (i.e., Sample 6) or contained some contaminating pixels (i.e., Samples 5 and 7). In such cases, due to the limited number of variables considered by the single-peak integration method, its performance was inferior to that of the other unsupervised method, leading to misclassification errors.

\subsection{Results of the similarity analysis based on the linear regression}

\paragraph{Background detection in the FTIR images of human thyroid and rat kidney tissue sections.} In the analysis of the FTIR image from tissue section samples, both the computed similarity score and residue, when arranged according to spatial information, effectively reflect the sample's contour and highlight the distinction between the sample and background regions. While both metrics capture the difference between sample and background, the similarity score shows relatively small variations among different background regions. In contrast, the similarity residue, serving as a complementary measure to the score, further reveals differences among background regions with distinct compositions—such as background spectra at the sample boundary versus those in distant background areas—thereby providing more detailed spatial and spectral information.

By tuning the thresholds of the similarity score and similarity residue, we can optimize the prediction results to be consistent with the labeled ground truth maps, which is viewed as the criterion in this task. However, due to the limited precision of the manual annotations, this method is capable of detecting background spectra within micro-pores in the sample—although these regions may not align with the annotations, they provide valuable additional information about background pixels, enabling the algorithm to meet other criteria in other tasks.

Moreover, compared to the integrated absorbance-based method, the range of integrated intensities within the same spectral band varies significantly between samples of different thicknesses (4 $\mu m$ for Samples 1–4 and 2 $\mu m$ for Samples 5–7), making it difficult to define a consistent and reliable threshold for classification with integrated absorbance. Furthermore, the minimum negative value in the integration results indicates that baseline drift in the integrated absorbance intensity can reach up to approximately -70 in some cases, which is non-negligible relative to the overall range of the integrated absorbance. Directly using cumulative absorbance for classification and detection of the background pixels would therefore lead to numerous misclassifications. In contrast, the proposed distribution-based approach, by defining a similarity score as the classification criterion, confines the measurement to a more appropriate and consistent range across different samples, thereby enhancing the consistency and practicality of the algorithm.

\begin{figure}[!htp]
    \centering
    \includegraphics[width=\linewidth]{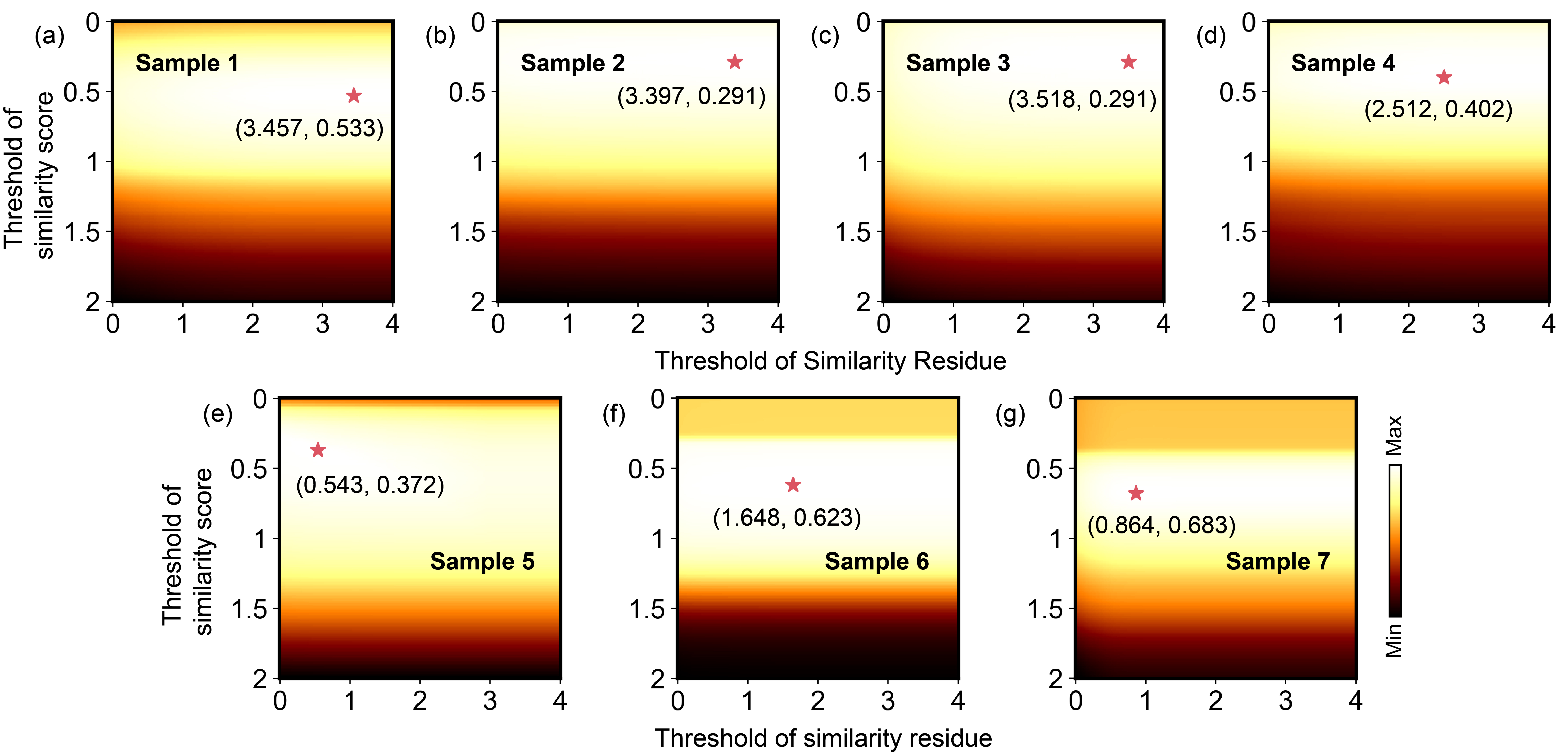}
    \caption{\textbf{Variation of the Jaccard index between the unsupervised approach predictions and the ground truth as a function of the similarity score and similarity residue thresholds.} The similarity score threshold is varied from 0 to 2, and the similarity residue threshold from 0 to 4. The resulting Jaccard index values across samples 1–7 are shown in the heatmaps (a–g). Additionally, the red stars on the plots indicate the threshold values corresponding to the maximum Jaccard index, which represent the optimal thresholds used for the results shown in Figure \ref{fig:UnsupervisedThyroid} and Figure \ref{fig:UnsupervisedKidney}.}
    \label{fig:Varying}
\end{figure}

\paragraph{Threshold varying analysis on the similarity score and the similarity residue.} To visually assess the impact of threshold selection on the results, we systematically traverse the threshold of the similarity score from 0 to 2 and the threshold of the similarity residue from 0 to 4. For each combination of thresholds, we predict the sample and background pixels for each sample. By computing the Jaccard index between the predicted segmentation maps and the ground truth, we generate a heatmap of the Jaccard index to visualize the performance across different threshold combinations, as shown in Figure \ref{fig:Varying}.

As shown in the figure, the Jaccard index first increases to a maximum and then decreases as the threshold of the similarity score varies from 0 to 2. In contrast, the Jaccard index remains relatively constant across the range of the thresholds of the similarity residue from 0 to 4. Therefore, in this method, the similarity score is the primary factor influencing the performance of background detection, while the similarity residue has a relatively minor impact. The residue threshold only plays a more significant role in samples with greater spectral variation and diverse background sources, such as samples 1, 3, and 7.

Therefore, this fully unsupervised background identification method detects background pixels based on the distributions of two informative parameters. The experiments have demonstrated that the selected parameters are highly correlated with the spectral distributions and provide sufficient information to distinguish background from sample pixels. Although the threshold selection offers considerable flexibility, the process itself introduces a degree of subjectivity that hinders full automation. Moreover, the coupling between the two parameters limits the method's overall freedom, potentially leading to misclassification of some contaminants as sample spectra. This limitation motivates the development of the following DNN-based automated background detection approach.

\subsection{Results of the automatic detection based on a deep neural network}

\begin{figure}[!htp]
    \centering
    \includegraphics[width=\linewidth]{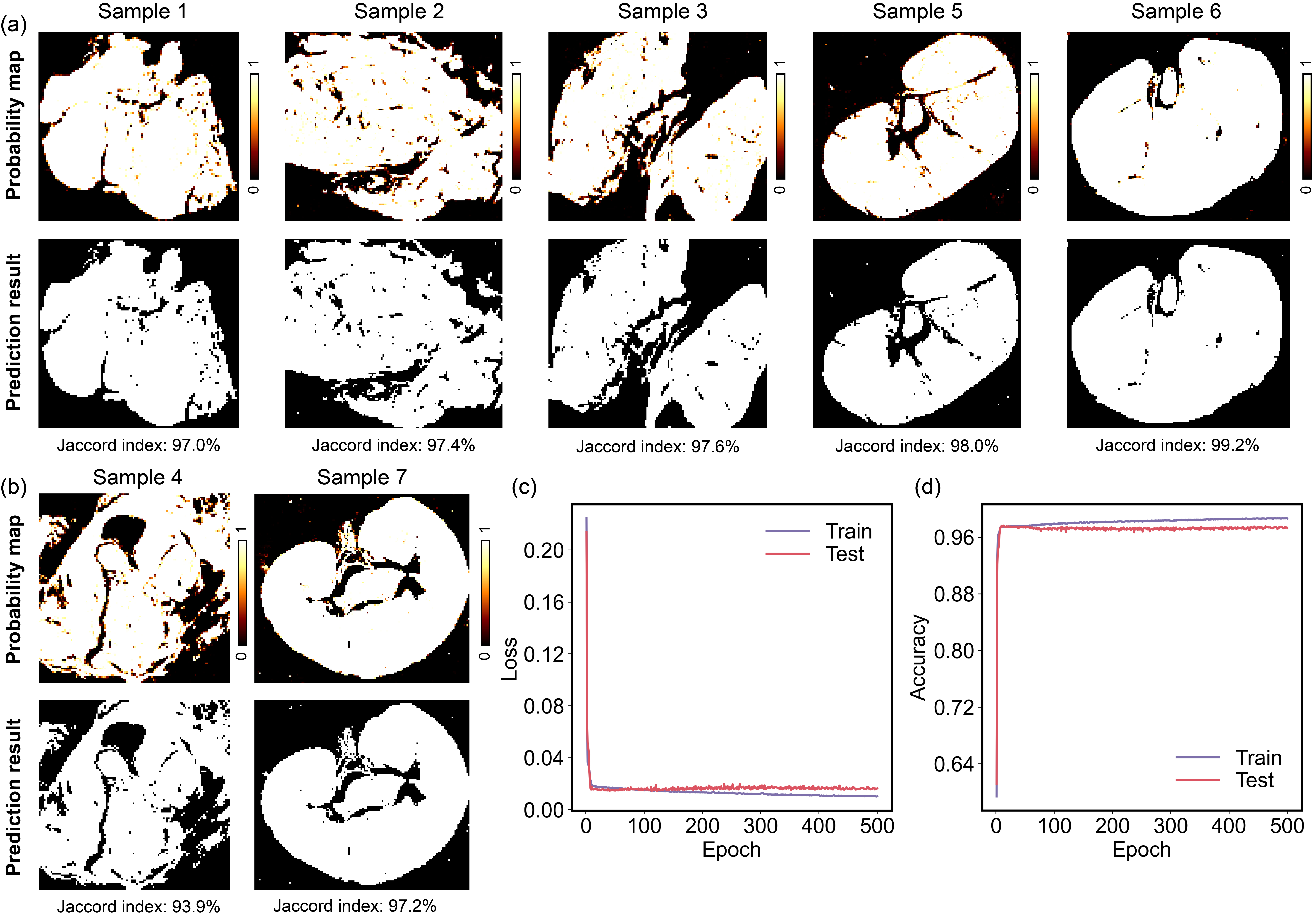}
    \caption{\textbf{Results of background detection with supervised deep neural network approach on tissue section samples.} For each sample, the probability map and the prediction result are shown. (a) The prediction results of the samples used for training. (b) The prediction results of the samples used for testing. (c) Mean loss after each epoch of training during the training procedure. (d) Mean prediction accuracy after each epoch of training during the training procedure.}
    \label{fig:dnnres}
\end{figure}

Since the DNN-based approach is a supervised method, we use the expected screening results shown in Figure \ref{fig:UnsupervisedThyroid} and \ref{fig:UnsupervisedKidney} as labels to generate data pairs for network training and testing. As mentioned earlier, we employed spectral data from 5 samples as the training set, comprising 76,597 spectra, and another 2 samples as the testing set, containing 30,632 spectra. Since the spectral data were shuffled during training, accuracy was chosen as the monitoring metric rather than the Jaccard index. The prediction loss and prediction accuracy on both the training and test sets exhibited consistent downward trends with minimal differences, demonstrating good convergence behavior, as shown in Figure \ref{fig:dnnres} c\&d.

As can be seen from the probability maps in Figure \ref{fig:dnnres} a\&b, the network predicts a probability less than 1 for edge pixels at the boundary between the sample and the background, indicating that the likelihood of these pixels belonging to the sample is not fully certain. This demonstrates that spectral features at the sample-background interface are partially coupled, which is consistent with common knowledge in conventional histopathological section preparation. From the prediction results, the network demonstrates strong predictive performance, exhibiting good fitting capability to the established criteria on both the training and test sets. Moreover, its strong performance on the test set indicates that the DNN possesses excellent generalization ability.

Compared to the proposed unsupervised method, although unsupervised approaches allow flexibility in parameter selection, and this flexibility introduces a degree of freedom in defining the criteria, the freedom can also hinder the overall automation process. In contrast, the DNN approach trains the network using fixed criteria, sacrificing some flexibility but enabling fully end-to-end automated prediction under predefined standards. Each method has its own advantages, and the choice between them should be determined based on the specific application context.

\section{Conclusion}

In this paper, we introduce the multivariate sample pixels identification algorithms for FTIR images implemented with unsupervised and supervised approaches. In the unsupervised methodology, besides the traditional integrated absorbance-based approach, we defined the similarity score and the similarity residue compared to the average spectrum of each sample IR image, calculated with a linear regression model. This new approach enables the detection of the sample pixels in images according to the distribution of the spectra and threshold selection, offering an effective and flexible pathway. In the supervised methodology, we proposed an end-to-end deep neural network model, enabling the automatic detection of the sample pixels. The model gives an accurate identification of the sample pixels and demonstrates good generalization capability. Based on the unsupervised and supervised frameworks, this paper comprehensively presents multiple different detection approaches, offering a thorough and robust solution to the problem. These findings have significant implications for FTIR signal processing and future research with FTIR images.

\section*{Ethical statement}

All experiments were performed in compliance with relevant laws and relevant institutional guidelines, and approved by the Ethics Committee of the First Hospital of China Medical University (Number 2024-582) and the Institutional Animal Care and Use Committee at Jiangsu Science Standard Medical Testing Company Limited (IACUC22-0107). Informed consent was obtained from human participants of this study as per the institute-approved standard protocol.

\section*{Disclosure statement}

The authors declare no conflicts of interest.

\section*{Acknowledgment}
The work is supported by the National Natural Science Foundation of China under Grant 62375170 and 62535019, the Shanghai Jiao Tong University under Grant YG2024QNA51, and the Science and Technology Commission of Shanghai Municipality under Grant 20DZ222040. We thank Z. Gui and Z. Wang, Department of Thyroid Surgery, the First Hospital of China Medical University, for providing human thyroid samples. We also thank Z. Qing, Shanghai General Hospital, Shanghai Jiao Tong University School of Medicine, for providing rat kidney samples.

\section*{Data availability}

Data underlying the results presented in this paper are not publicly available at this time but may be obtained from the authors upon reasonable request.

\bibliographystyle{unsrt}  
%\bibliography{references}  %%% Remove comment to use the external .bib file (using bibtex).
%%% and comment out the ``thebibliography'' section.

%%% Comment out this section when you \bibliography{references} is enabled.
\bibliography{references}

\end{document}